\documentstyle[preprint,eqsecnum,aps]{revtex}

\begin{document}

\draft

%======================================%
%<<<<<<<<<<<< TITLE PAGE >>>>>>>>>>>>>>%
%======================================%

\preprint{YITP-97-63, gr-qc/9712018}

\title{
Thermodynamics of entanglement in Schwarzschild spacetime
}

\author{
Shinji Mukohyama$^\dagger$,  
Masafumi Seriu$^\ddagger$
and 
Hideo Kodama$^\dagger$
}

\address{
${}^\dagger$Yukawa Institute for Theoretical Physics \\
Kyoto University, Kyoto 606-01, Japan \\
and \\
${}^\ddagger$ Physics Group, Department of Education \\
Fukui University, Fukui 910, Japan
}

\date{\today}

\maketitle

%======================================%
%<<<<<<<<<<<<< ABSTRACT >>>>>>>>>>>>>>>% 
%======================================%

\begin{abstract}

Extending the analysis in our previous paper, we construct the
entanglement thermodynamics for a massless scalar field on the
Schwarzschild spacetime. Contrary to the flat case, the entanglement
energy $E_{ent}$ turns out to be proportional to area radius of the
boundary if 
it is near the horizon. This peculiar behavior of $E_{ent}$ can be
understood by the red-shift effect caused by the curved background.
Combined with the behavior of the entanglement entropy, this result
yields, quite surprisingly, the entanglement thermodynamics of the
same structure as the black hole thermodynamics.  On the basis of
these results, we discuss the relevance of the concept of entanglement
as the microscopic origin of the black hole thermodynamics.

\end{abstract}


\newpage

%%%%%%%%%%%%%%%%%%%%%%%%%%%%%%%%%%%%%%%%%%%%%%%%%%%%%%%%%%%%%%%%%%%%
%%%%%%%%%%%%%%%%%%%%%%%%%%%%%%%%%%%%%%%%%%%%%%%%%%%%%%%%%%%%%%%%%%%%
%SECTION 1 
%%%%%%%%%%%%%%%%%%%%%%%%%%%%%%%%%%%%%%%%%%%%%%%%%%%%%%%%%%%%%%%%%%%%
%%%%%%%%%%%%%%%%%%%%%%%%%%%%%%%%%%%%%%%%%%%%%%%%%%%%%%%%%%%%%%%%%%%%

\section{Introduction}
\label{sec:Introduction}

One of the central problems in black hole physics is the 
identification of the microscopic origin of  black hole entropy, 
which obeys the relation 
%============< EQUATION >==============%
%
\begin{equation}
 S_{BH} = \frac{1}{4 l_{Pl}^2}{\cal A},	
\label{eqn:S-BH}
\end{equation}
%======================================%
where $\cal A$ is the area of the event horizon and 
$l_{\rm Pl}:=\sqrt{G}$ is the Planck 
length~\cite{WALD}. (Hereafter we will set $c=\hbar=1$.) 
There are two important facts which suggest that the black hole entropy
may have some microscopic origin. One is Hawking's 
argument~\cite{Hawking1975} on quantum fields in a black hole spacetime 
showing that the black hole emits thermal radiation with the temperature 
$T_{BH}$ which is proportional to that determined by the classical 1st 
law for black holes \cite{Bekenstein} with a universal coefficient. In 
fact, the coefficient on the right-hand side of Eq.(\ref{eqn:S-BH}) is 
chosen so that $T_{BH}$ coincides with the latter. For derivation of 
$T_{BH}$, see \cite{Israel,Laflamme} as well as
\cite{Hawking1975}. The other is the fact 
that the free energy calculated from the Euclidean path integral for the 
pure gravity system gives in the saddle point approximation exactly the 
same expressions for the temperature and the entropy as those given 
above. On the basis of these facts and their consistency, there have been 
proposed various candidates for the microscopic origin of the black hole 
entropy~\cite{BHentropy}.
Among them the simplest one is the idea of the entanglement 
entropy\cite{BKLS,Sred}, which focuses on the entropy associated with
states of quantum fields on black hole spacetimes and is strongly 
motivated by the first of the above two facts.

The entanglement entropy   itself is quite a general 
concept which is nothing but coarse graining entropy for a
quantum system caused by an observer's partial ignorance of 
the information on the  state. Now the idea is that 
the event horizon may  play the role of the boundary of perception
 for an asymptotic observer so that  
its   existence may give rise to the  entropy. 
Indeed for simple models on a flat spacetime\cite{BKLS,Sred}
explicit estimations showed that the entanglement entropy $S_{ent}$ 
is always proportional to the area ${\cal A}$ of 
the  boundary between two regions of a spatial section of spacetime:
%============< EQUATION >==============%
%
\begin{equation}
S_{ent} ={C_S \over a^2}{\cal A},
\label{eqn:SproptoA}
\end{equation}
%======================================%
where $a$ is a cut-off scale for regularization and $C_S$ is a model-dependent
coefficient of order unity.
Thus there is  a clear similarity between $S_{ent}$ and  $S_{BH}$. 

In order to see whether there is something deeper in this similarity,
we have constructed a kind of thermodynamics of a space boundary, 
which we call entanglement thermodynamics, for a massless scalar field 
on flat spacetime in our previous paper\cite{paper1}. There,
by giving a suitable definition of the entanglement energy $E_{ent}$, 
we determined the entanglement temperature $T_{ent}$ by imposing
the first law of thermodynamics.  Then we compared the system
of thermodynamics quantities $(E_{ent}, S_{ent}, T_{ent})$ obtained
by this procedure with the corresponding one for black hole,
$(E_{BH}=M, S_{BH}, T_{BH})$. This comparison showed that 
the entanglement thermodynamics on a flat background possesses 
a totally different structure  compared with the black hole one.

It is not difficult to  understood why such discrepancy occurs.
Since we wanted to construct thermodynamics of a space boundary, 
we have defined the entanglement energy $E_{ent}$ so that
it depends only on quantum degree of freedom around the space boundary.
As a result, $E_{ent}$ became proportional to the boundary area 
${\cal A}$ unlike the black hole energy $E_{BH}=M\propto \sqrt{\cal A}$.

Since the dependence of the thermodynamical quantities on the boundary 
area come from the very nature of the idea of entanglement, it may appear
that the idea of entanglement cannot have any relevance to the black hole
thermodynamics beyond the similarity of the expressions for entropy. 
However, it is not the case. It is because gravity is not taken into account
in the above argument. In fact there is a good reason to expect that
the inclusion of gravity improves the situation drastically.

In black hole spacetime the energy of quantum field gets gravitational
corrections, which depend on the definition of the energy. Taking into 
account of the fact that $E_{ent}$ depends only on modes around the
boundary, the entanglement energy estimated on flat background, 
$E_{ent,{\rm flat}}$, should be identified with the energy 
$E_{ent, {\rm b}}$ measured by
a Killing observer located near the horizon. On the other hand,
since the black hole mass used as the energy in the black hole 
thermodynamics corresponds to the energy measured by an observer
at spatial infinity, it is  natural to use the corresponding
quantity $E_{ent,\infty}$. Due to the gravitational redshift,
these energies are related by
%============< EQUATION >==============%
%
\begin{equation}
E_{ent,\infty}= (-g_{tt})_{\rm b}^{1/2}E_{ent, {\rm b}}
\simeq (-g_{tt})_{\rm b}^{1/2} E_{ent,{\rm flat}},
\end{equation}
%======================================%
where $(-g_{tt})^{1/2}_{\rm b}$ is the well-known red-shift 
factor\cite{WALD} for a signal emitted at the boundary. 

Let us assume that the boundary is at the proper distance $a\sim l_{\rm Pl}$ 
from the horizon. Then for the Schwarzschild metric we obtain 
$(-g_{tt})^{1/2}_{\rm b}\simeq a/(2M)\propto 1/\sqrt{{\cal A}}$. Hence from
$E_{ent,{\rm flat}} \propto {\cal A}$, if follows that
$E_{ent,\infty} \propto \sqrt{\cal A}$.
On the other hand, $S_{ent}$ is independent of the position of 
an observer once a quantum state is fixed, so  $S_{ent} \propto {\cal A}$ 
as before. Thus it is expected that the gravity effect modifies the
structure of the entanglement thermodynamics so that it becomes identical
to that of the black hole thermodynamics. 

On the basis of these observations, in this paper we construct
the entanglement thermodynamics for a massless scalar field on
the Schwarzschild background, and compare its structure with that 
of the black hole thermodynamics. 

The paper is organized as follows. In the next section we describe the
model used in the paper in detail, and clarify the definitions of the
entanglement entropy and energy and  the basic assumptions with brief
explanations of their motivations. Then, after explaining our
regularization scheme, we derive general formulas for the
entanglement entropy and energy of the regularized system and estimate 
them numerically in \S3 and \S4. On the basis of these results, we
compare the structure of the entanglement thermodynamics for the 
massless scalar field in the Schwarzschild background with that of 
the black hole in \S5.

%%%%%%%%%%%%%%%%%%%%%%%%%%%%%%%%%%%%%%%%%%%%%%%%%%%%%%%%%%%%%%%%%%
%%%%%%%%%%%%%%%%%%%%%%%%%%%%%%%%%%%%%%%%%%%%%%%%%%%%%%%%%%%%%%%%%%
%SECTION 2
%%%%%%%%%%%%%%%%%%%%%%%%%%%%%%%%%%%%%%%%%%%%%%%%%%%%%%%%%%%%%%%%%%
%%%%%%%%%%%%%%%%%%%%%%%%%%%%%%%%%%%%%%%%%%%%%%%%%%%%%%%%%%%%%%%%%%
\section{Model construction}
\label{sec:model}

In order to construct the entanglement thermodynamics of a black hole,
we must make clear how to implement the idea of entanglement into a
quantum field system on a black hole spacetime, and give the
definitions of the basic thermodynamical quantities in terms of the
quantum fields.

\subsection{Model description}

The basic idea of the entanglement thermodynamics is to express
the thermodynamical quantities for a black hole in terms of 
expectation values of quantum operators dependent on the
spacetime division as in the statistical mechanics modeling
of the thermodynamics for ordinary systems. Therefore we must
specify how to divide spacetime into two regions and with respect
to what kind of state the expectation values are taken. 

According to the original idea of entanglement, it is clearly most
natural to consider a dynamical spacetime describing black hole
formation from a nearly flat spacetime in the past infinity, and
divide the spacetime into the regions inside and outside the
horizon. In this situation, if we start from the asymptotic Minkowski
vacuum in the past, the entanglement entropy associated with the
division of spacetime by the horizon acquires a clear physical meaning.
However, this ideal modeling cause difficulties.  The most serious one
is caused by the occurrence of the Hawking radiation: its contribution
to the entanglement entropy diverges. Of course, if the backreaction
effect is properly taken into account, this divergence may
disappear. But such a modeling is intractable at present and is
beyond the scope of the present paper. 

In order to avoid this difficulty, we replace the problem by a
stationary one on the basis of the following arguments. First,
consider a situation in which a thin spherical shell with a tiny mass
$\delta M$ infalls toward the origin to form a Schwarzschild black
hole with mass $\delta M$. In this situation, the timelike Killing
surface $\Sigma$ with $r=2G\Delta M$ in the Minkowski region turns to
the horizon after the passage of the shell. Two things happen in this
process: a division of spacetime by the horizon and a change of
geometry.  Among these two things, the latter is responsible for the
Hawking radition, because the Hawking radiation is essentially
determined by the geometry near the horizon in the geometrical optics 
approximation. Meanwhile, the division of spacetime produces the
entropy of the black hole in our approach based on the
entanglement. Thus, we can determine the entropy of the black hole
avoiding the above difficulty caused by the Hawking radition, if we
calculate the entanglement entropy associated with the division of
Minkowski spacetime by the timelike surface $\Sigma$ starting from the
Minkowski vacuum. Because we need to introduce a cut-off length $a$ of
the order of the Planck length $l_{Pl}$, a natural minimum value of
$\Delta M$ is determined by the condition $r^2\sim a^2$. The
corresponding value of the entropy becomes of order unity.

Next, let us consider a thin spherical shell with a tiny mass $\Delta
M$ infalling toward the center in a Schwarzschild background spacetime
with mass $M$. After the passege of the mass shell, a Schwarzschild
black hole with mass $M+\Delta M$ is formed. As in the previous case,
this process produces a new horizon which divides the spacetime by a
null surface corresponding to the stationary timelike surface $\Sigma$
with $r=2G(M+\delta M)$ before the passege of the shell, and slightly
modifies the geometry. If we apply the reasoning in the previous case
to the present situation, we can expect that the entropy of the new
black hole is given by the entanglement entropy for quantum fields in
the Killing vacuum state, associated with the division of the
Schwarzschild spacetime by the timelike surface $\Sigma$ before the
passage of the shell. Here, the Killing vacuum state should be taken
in order to avoid the Hawking radiation. Now the minimum value of
$\Delta M$ is determined by the condition that the increase of the
area of the horizon is of order $a^2$, which corresponds to an
increase of entropy of order unity. For this minimum value, the proper
distance of $\Sigma$ to the original horizon becomes of order $a$ if
$a\sim\l_{Pl}$.

On the basis of these arguments, in this paper, we consider a quantum
field in the Killing vacuum state in the Schwarzschild background with 
mass $M$,
%============< EQUATION >==============%
%
\begin{equation}
	ds^2 = -(1-r_{0}/r)dt^2 + (1-r_{0}/r)^{-1}dr^2 
			+ r^2 d\Omega^2,
\label{eqn:usualmetric}
\end{equation}
%======================================%
where $r_0=2GM$, and calculate the entanglement entropy and the
entanglement energy to be defined later associated with the division
of the spacetime by a timelike surface $\Sigma$ at a proper distance
of the order of a cut-off length($\sim l_{Pl}$) to the horizon. 
Since there is no definite criterion on the exact position of the
boundary in our framework, we will also investigate the influence of
the variation of its position.

Here a subtlety occurs: if we require that the quantum field is in
the Killing vacuum state on the whole extended Kruscal spacetime and
take the horizon as the boundary surface, then the
entanglement entropy vanishes because the state is expressed as the
tensor product of the Killing states in the regions I and II (
{\it Figure} \ref{fig:Kruskal}). This
triviality comes from the fact that the quantum degree of freedom
on the horizon is completely eliminated for such a state. In order 
to avoid this, we restrict the spacetime into the 
region I, and replace the boundary by a time-like surface $\Sigma$ at
a proper distance of the order of the cut-off length (typically, 
Planck length) of the theory to the horizon~\footnote{
Our present model clears such a criticism~\cite{Frolov&Novikov} for 
works done before~\cite{BKLS,Sred,paper1,Sent} that in these works the
boundary had always been chosen as a timelike surface, contrary to the
event horizon, which is a null surface.
}. 
The boundary $\Sigma$ can be regarded as a would-be horizon when 
matter fields (a real scalar field in our analysis in this paper), 
which contribute to the entanglement entropy, falls into the black hole. 
Evidently, the would-be horizon must be very close to the horizon 
since our analysis is semi-classical one. However, because of the 
existence of the cut-off length there is the minimum proper length 
between the would-be and the present horizon: it must be of order of 
the cut-off length. Hence it is natural to take as the boundary a 
timelike surface at a proper distance of the order of the cut-off length 
to the horizon. This prescription is expected to give a correct 
estimate of the thermodynamical quantities for the case in which the 
spacetime is divided by the horizon and the quantum state inside the horizon 
is arbitrary specified because the entanglement quantities depend only on
the degrees of freedom within the distance of the order of the cut-off
length from the boundary. However, we should keep in mind that we have
no definite criterion regarding 
the exact position of the boundary. To minimize this ambiguity, we will 
also investigate the influence of the variation of the boundary position.

As a matter content we consider a real scalar field described by
%============< EQUATION >==============%
%
\begin{equation}
  S = -\frac{1}{2}\int 
  \{ \partial^{\mu}\phi \partial_{\mu}\phi + \mu^2 \}\sqrt{-g} dx^4\ \ .
\label{eqn:action}
\end{equation}
%======================================%
The mass $\mu$ does not play an essential role since a typical length
scale controlling the entanglement thermodynamics is much smaller than
the Compton length of a usual field.  Therefore we just set $\mu=0$
in the numerical computations.

\subsection{Entanglement entropy and energy}\label{subsec:energy}

There is no difficulty in generalizing the entanglement entropy for 
the flat case\cite{BKLS,Sred,paper1} to the case of Schwarzshild 
background. In contrast, with regards to the definition of
the entanglement energy, we cannot simply extend the definition
used for the flat case\cite{paper1} due to the difference in the method of
implementation of the idea of entanglement as well as due to the
existence of the black hole mass in the present case. 

Before discussing the definition of the entanglement energy, let us
first recall the standard general procedure of building a statistical
mechanics model for thermodynamics.  Let $A$ be a subsystem surrounded
by the environment $B$.  For a quasi-static process, the first law of
thermodynamics for the subsystem $A$ is expressed as
%============< EQUATION >==============%
%
\begin{equation}
dE=TdS + fdx\ \ .
\label{eqn:1stlaw}
\end{equation}
%======================================%
Here $E$, $S$ and $T$ are the thermodynamical energy, entropy and
temperature of $A$, respectively, $f$ is a generalized force applied
to $A$, and $x$ is a generalized displacement.  Now the statistical
mechanics model of this thermodynamics is constructed by the following
sequence of steps:
\begin{description}
      \item[(a)] Assign a microscopic model to $A$ and construct 
                 a Hamiltonian system for it.
      \item[(b)] Make statistical assumptions  
                  (e.g. the principle of equi-weight) which 
                  define  averaging procedures for the microscopic 
                  variables.
      \item[(c)] Identify  the thermodynamical variables  with 
            quantities constructed from the microscopic  
            variables introduced in  $(a)$ along with the 
             averaging procedure in   $(b)$.
      \item[(d)] Reproduce  the  relations between the 
           thermodynamical variables (thermodynamical 
           laws in particular).
\end{description}

Since the basic idea of the entanglement thermodynamics is to 
construct a statistical mechanics model of the black hole 
thermodynamics using the idea of entanglement, we should also
follow a similar procedure to this general one in defining
$E_{ent}$. 

In our case it is natural to regard the scalar field in the region
inside the boundary $\Sigma$ as the subsystem $A$, and the scalar
field in the region outside $\Sigma$ as the environment
$B$. Then, if we decompose the Hamiltonian $H$ corresponding to 
the Killing energy to the parts dependent on the microscopic degrees
of freedom inside $\Sigma$, those outside $\Sigma$ and both as
$H=H_{in}+H_{out}+H_{int}$,  the most 
natural choice of the microscopic variable for $E_{ent}$ is the
operator $H_{in}$. Clearly the averaging procedure is given by
taking the expectation value with respect to the Killing vacuum
of the total Hamiltonian $H$. Hence we are led to the
definition
\begin{description}
\item[(I)]   $E_{ent}^{(I)}=\langle:H_{in}:\rangle$,
\end{description}
where $:H_{in}:$ denotes the normal ordering with respect to
the ground state of $H_{in}$. 

Here note that the choice
\begin{description}
\item[(I')]   $E_{ent}^{(I')}=\langle:H_{out}:\rangle$
\end{description}
is essentially same as $(I)$ in spite of its  appearance,
as we will see in Sec.\ref{sec:energy}.
It is because $E_{ent}$ measures a sort of  
disturbance to the vacuum state caused by the boundary $\Sigma$, 
and its value is determined just by the modes around $\Sigma$.
Hence the difference between $(I)$ and $(I')$
comes from   a tiny  difference in the redshift effect on 
the modes just inside $\Sigma$ and just outside $\Sigma$.

These choices should be contrasted with the definition of the energy
in the standard argument of black hole thermodynamics. There 
the energy is given by the black hole mass $M$, which is
quite natural in the framework in which the backreaction of matter
on gravity is consistently taken into account. However, in the
present case, the option
\begin{description}
\item[(II)]   $E_{ent}^{(II)}=M$,
\end{description}
lacks charm in our framework because it is not related to 
microscopic degrees of freedom. Of course, it may be reasonable to
replace the definition (I) by
\begin{description}
\item[(III)]   $E_{ent}^{(III)}=M+\langle:H_{in}:\rangle$,
\end{description}
in the sense that it represents a kind of total energy of the
system consisting of the scalar field and the gravitational
field.

Here we should comment on a subtlety of the role of $M$ in our
model. Since we are neglecting the backreaction of the matter on
gravity, the black hole mass $M$ determining the background
geometry $g_{ab}$ might be regarded as the external parameter $x$
in the above expression for the 1st law of thermodynamics. 
However, since the Killing vacuum state depends on $M$ as we will see
later,  $M$ becomes a function of thermodynamical variables 
$S_{ent}$ and $E_{ent}$. Thus in effect the term $fdx$ 
in Eq.(\ref{eqn:1stlaw}) does not appear in the present 
case\cite{Frolov.V1995}.\footnote{ It is not a surprise that 
an external parameter becomes a function of variables
describing thermodynamics state. For example, in the Rindler
spacetime, the acceleration $\alpha$ is determined by the 
Rindler temperature.}

Because there is no firm ground to pick up one of these, we
will investigate all of them. Further, for the sake of 
comparison with the result in the flat spacetime models
discussed in our previous paper\cite{paper1}, we also calculate
the quantities defined by
\begin{description}
\item[(${\rm IV}_1$)]   $E_{ent}^{(IV_1)}:=\langle:H:\rangle_{\rho'}$, 
\item[(${\rm IV}_2$)]   $E_{ent}^{(IV_2)}=\langle:H_{in}:\rangle
                   +\langle:H_{out}:\rangle$.
\end{description}
Here $:H:$ and $:H_{out}$ denote the normal products with respect to
the ground states of $H$ and $H_{out}$, respectively, and
$\langle\cdot\rangle_{\rho'}$ indicates the expectation value with
respect to the density matrix $\rho' :=\rho_{in} \otimes \rho_{out}$,
where $\rho_{in}$ and $\rho_{out}$ are the reduced density matrices
for the inside and the outside of $\Sigma$, respectively\cite{paper1}.

Finally we comment on the regularization. In dealing with the matter
field, we naturally encounter the ultraviolet divergence. We adopt
here the cut-off regularization by introducing the length scale $a$,
which is supposed to be of order of $l_{\rm Pl}$. On dimensional
grounds we expect that $S_{ent} \propto {\cal A} /a^2$ and $E_{ent}
\propto \sqrt{A}/a^2$. Thus the entanglement temperature
$T_{ent}=dE_{ent}/dS_{ent}$ is expected to be independent of the
cut-off scale $a$.  We will see later that this is the case except for
the definitions $(II)$ and $(III)$.

%%%%%%%%%%%%%%%%%%%%%%%%%%%%%%%%%%%%%%%%%%%%%%%%%%%%%%%%%%%%%%%%%
%%%%%%%%%%%%%%%%%%%%%%%%%%%%%%%%%%%%%%%%%%%%%%%%%%%%%%%%%%%%%%%%%
%SECTION 3
%%%%%%%%%%%%%%%%%%%%%%%%%%%%%%%%%%%%%%%%%%%%%%%%%%%%%%%%%%%%%%%%%
%%%%%%%%%%%%%%%%%%%%%%%%%%%%%%%%%%%%%%%%%%%%%%%%%%%%%%%%%%%%%%%%%

\section{Entanglement entropy}
\label{sec:entropy}

\subsection{Basic formulas}
\label{subsec:basic} 
 
In order to estimate the entanglement entropy and the entanglement
energy for the scalar field on the Schwarzschild spacetime, we first
regularize the field theory described by Eq.(\ref{eqn:action}) and
reduce it to a discrete canonical system described by a Hamiltonian of the
following form :
%============< EQUATION >==============%
%
\begin{equation}
 	H_{0} = \sum_{A,B=1}^N 
 	        \frac{1}{2a}\delta^{AB}p_{A}p_{B} 
 		+ \frac{1}{2}V_{AB}q^Aq^B, 
\label{eqn:Hamiltonian}
\end{equation}
%======================================%
where $\{ ( q^A, p_A ) \}$ $(A=1,2,\cdots,N)$ are canonical pairs.
Here both $q^A$'s and $p_A$'s are of physical dimension $[L^0]$, and
the parameter $a$ is the cut-off length of order of $l_{\rm Pl}$.

Because of the spherical symmetry of the system, if we expand the
scalar field $\phi$ in terms of the real spherical harmonics as
%============< EQUATION >==============%
%
\begin{equation}
	\phi (t, \rho, \theta,\varphi) 
	= \sum_{l,m}\phi_{lm}(t,\rho)Z_{lm}(\theta,\varphi),
\label{eqn:expansion}
\end{equation}
%======================================%
the action becomes a simple sum of the contributions from each
mode $\phi_{lm}$. 
Here $Z_{lm}=\sqrt{2}\Re Y_{lm}$, $\sqrt{2}\Im Y_{lm}$ for $m>0$ and 
$m<0$, respectively, with $Z_{l0}=Y_{l0}$, and  $\rho$ is a suitable 
radial coordinate. Hence if we discretize this radial coordinate, we 
obtain a regularized system.

As the radial coordinate, we adopt the proper length from the horizon,
which is related to the circumferential radius coordinate $r$ by
%============< EQUATION >==============%
%
 \begin{equation}
 \rho =\int_{r_0}^r \frac{dr}{\sqrt{1-\frac{r_0}{r}}} 
      =r_0 \left\{ \frac{\nu}{1-\nu^2} 
           + \ln \frac{1+\nu}{\sqrt{1-\nu^2}} \right\}, 
 \label{eqn:rho}
 \end{equation}
%======================================%
where $\nu:=\sqrt{1-\frac{r_0}{r}}$. In this coordinate the metric 
Eq.(\ref{eqn:usualmetric}) is written as
%============< EQUATION >==============%
%
\begin{equation}
	ds^2 = -\nu^2 dt^2 + d\rho^2 
			+ r_0^2 \frac{d\Omega^2}{(1-\nu^2)^2},
\label{eqn:newmetric}
\end{equation}
%======================================%
where $\nu$ is understood as a function of $\rho$ through 
Eq.(\ref{eqn:rho}).

Plugging Eqs.(\ref{eqn:newmetric}) and (\ref{eqn:expansion}) into
Eq.(\ref{eqn:action}), we get
%============< EQUATION >==============%
%
\begin{equation}
S=\frac{1}{2}\sum_{lm}\int dt\ d\rho\ 
   \left[   \frac{r_0^2 \dot{\phi}_{lm}^2}{\nu (1-\nu^2)^2} 
         -\left\{ 
                             \frac{r_0^2 \nu}{(1-\nu^2)^2} 
                         \left(\partial_\rho \phi_{lm} \right)^2   
                         + \left( l(l+1) + \frac{(r_0 \mu)^2}{(1-\nu^2)^2}\right)
                           \nu \phi_{lm}^2 
                      \right\}       
   \right].
\end{equation}
%======================================%
Comparing this equation with Eq.(\ref{eqn:Hamiltonian}) we see that
the most suitable configuration variable is the dimensionless
one defined by
%============< EQUATION >==============%
%
\begin{equation}
 \psi_{lm}(t,\rho):=\frac{r_0}{\nu^{1/2}(1-\nu^2)} 
                     \phi_{lm}(t,\rho).
\end{equation}
%======================================%
Hereafter we adopt a boundary condition that 
$\psi_{lm}(t,\rho)$ is finite at the horizon, which 
corresponds to the Dirichlet boundary condition for the original variable 
$\phi_{lm}(t,\rho)$. To be precise, in defining Hamiltonian for our system, 
which must be self-adjoint, there are two options for the boundary 
condition of $\phi_{lm}(t,\rho)$ at the horizon: the Dirichlet boundary 
condition and the Neumann boundary condition. In our case we adopt the 
former since the later makes the Killing energy divergent. 

Now it is straightforward to  perform  the canonical 
transformation. The Poisson bracket relations become
%============< EQUATION >==============%
%
\begin{eqnarray}
        \{ \psi_{lm}(\rho), \pi_{l'm'}(\rho')\}
         & = & \delta_{ll'}\delta_{mm'}\delta(\rho -\rho') \ \ ,\nonumber\\
         \{  \psi_{lm}(\rho), \psi_{l'm'}(\rho')\}
         & = & 0 \ \ ,\nonumber\\
         \{ \pi_{lm}(\rho), \pi_{l'm'}(\rho')\}
         & = & 0 \ \ ,
\end{eqnarray}
%======================================%
where $\pi_{lm}$ is a momentum conjugate to $\psi_{lm}$ with  
dimension  $[L^{-1}]$. In terms of these canonical quantities
the Hamiltonian is expressed as
%============< EQUATION >==============%
%
\begin{eqnarray}
        H & =& \sum_{lm}H_{lm}\ \ ,   \nonumber \\
        & H_{lm} &=\frac{1}{2}\int d\rho\pi_{lm}^2(\rho)
              + \frac{1}{2}\int d\rho d\rho'
              \psi_{lm}(\rho)V_{lm}(\rho,\rho')\psi_{lm}(\rho'),
\label{eqn:Hamiltonian1}
\end{eqnarray}
%======================================%
where
%============< EQUATION >==============%
%
\begin{eqnarray}
              \psi_{lm}(\rho)V_{lm}(\rho,\rho')\psi_{lm}(\rho')
         &=&\delta (\rho,\rho')
         \Big[
           \frac{\nu}{(1-\nu^2)^2} 
                 \Big\{ 
                     \partial_\rho 
                      \left( \nu^{1/2}(1-\nu^2)\psi_{lm} \right) 
                  \Big\}^2 \nonumber \\
           &{}& +\frac{\nu^2}{r_0^2} 
              \left\{ l(l+1)(1-\nu^2)^2 + (r_0 \mu)^2
                 \right\} \psi_{lm}^2  
         \Big].
\label{eqn:potential}
\end{eqnarray}
%======================================%

We regularize this system by replacing it by a difference system with
respect to $\rho$ with spacing $a$. To be precise, we make the
following replacements:
%============< EQUATION >==============%
%
\begin{eqnarray*}
\rho \qquad \qquad   & \to & Aa, \\
\psi_{lm}(\rho=(A-1/2)a) & \to & q_{lm}^A, \\
\pi_{lm} (\rho=(A-1/2)a) & \to & p_{lmA}, \\
\nu (\rho=(A-1/2)a) & \to & \nu_A, \\
\delta(\rho=Aa, \rho=Ba) & \to & \delta_{AB}/a,
\end{eqnarray*}
%======================================%
where $A$ runs over the positive integers. To achieve a better
precision, we adopt the middle-point prescription in discretizing the
terms including a derivative: we replace a term, say,
$f(\rho)\partial_\rho g(\rho)$ by $f_{A+1/2} \cdot \frac{1}{a}
(g_{A+1}-g_A)$. Further, in order to make the degrees of the system
finite, we impose the boundary conditions $q_{lm}^{N+1}=0$.
In the numerical calculation $N$ is taken to be sufficiently large
so that this artificial boundary condition, which is required just for a
technical reason, does not affect the results.

In this manner we get the hamiltonian in the desired form:
%============< EQUATION >==============%
%
\begin{eqnarray}
        H_0 & =& \sum_{lm}H_0^{(lm)}\ \ ,   \nonumber \\
 	H_0^{(lm)} & = & \sum_{A,B=1}^N \left[
 	        \frac{1}{2a}\delta^{AB}p_{lmA}p_{lmB} 
 		+ \frac{1}{2}V_{AB}^{(lm)}q_{lm}^A q_{lm}^B \right]\ \ , 
\label{eqn:Hamiltonian2}
\end{eqnarray}
%======================================%
where
%============< EQUATION >==============%
%
\begin{eqnarray*}
      \sum_{A,B=1}^N V_{AB}^{(lm)}q_{lm}^A q_{lm}^B 
       & = & \frac{1}{a}\sum_{A=1}^N \left[
         \frac{\nu_{A+1/2}}{(1-\nu_{A+1/2}^2)^2}
         \left(
              \nu_{A+1}^{1/2}(1-\nu_{A+1}^2)q_{lm}^{A+1}  -
              \nu_A^{1/2}(1-\nu_A^2)q_{lm}^A \right)^2 
              \right.  \\
          & +& \left. \left( \frac{a}{r_0} \right)^2 \nu_A^2
               \left(l(l+1)(1-\nu_A^2)^2 + (r_0 \mu )^2 \right) 
               { q_{lm}^A }^2  \right]. 
\end{eqnarray*}
%======================================%
Here $V^{(lm)}$ becomes   the  positive definite, symmetric matrix
 whose explicit form is   
%============< EQUATION >==============%
%
\begin{eqnarray}
 \left( V^{(lm)}_{AB}\right) & = & \frac{2a}{r_0^2}\left( 
               \begin{array}{cccccc}
                   \Sigma^{(l)}_1 & \Delta_1 & & & & \\
         \Delta_1 & \Sigma^{(l)}_2 & \Delta_2 & & & \\
       & \ddots & \ddots & \ddots & & \\
       & & \Delta_{A-1} & \Sigma^{(l)}_A & \Delta_A & \\
       & & & \ddots & \ddots & \ddots 
               \end{array}
             \right) \ \ \ , \nonumber\\
   \Sigma^{(l)}_A
     & = & \frac{r_0^2}{2a^2} \nu_A (1-\nu_A^2)^2\left(
          \frac{\nu_{A+1/2}}{(1-\nu_{A+1/2}^2)^2}
          + \frac{\nu_{A-1/2}}{(1-\nu_{A-1/2}^2)^2}
            \right) \nonumber\\
     &   &  + \frac{1}{2} \nu_A^2  
            \left(l(l+1)(1-\nu_A^2)^2 + (r_0 \mu)^2 \right),\nonumber\\
   \Delta_A
      & = & - \frac{r_0^2}{2a^2} \nu_{A+1/2}(\nu_{A+1}\nu_A)^{1/2}
         \frac{(1-\nu_{A+1}^2)(1-\nu_A^2)}{(1-\nu_{A+1/2}^2)^2 }\ \ .
\label{eqn:Vlm}
\end{eqnarray}
%======================================%

Here note that the mass term in $\Sigma^{(l)}_A$ is negligible
compared with the first term if $a\mu << 1$.  Therefore we will simply
set $\mu=0$ in the numerical calculations.

\subsection{Formulas for the entanglement entropy}

In the discretized system\footnote{ For the notational simplicity we
will often omit the suffices $(l,m)$ if no confusion occurs.}  $\left
( (q^A,p_A);H_0\right)$ $(A=1,2,\cdots,N)$ with
Eq.(\ref{eqn:Hamiltonian2}), the position of the space boundary
$\Sigma$ is given by $\rho=\rho_B:=n_Ba$ with $n_B =O(1)$, which
the set of modes $\{q^A\}$ $(A=1,\cdots,N)$ into two subsets, $\{ q^a\}$
$(a=1,\cdots,n_B)$ and $\{ q^{\alpha}\}$ $(\alpha =n_B+1,\cdots,N)$.
Here we regard $\{ q^a\}$ and $\{ q^\alpha \}$ as the inside modes and
the outside modes, respectively. 

The Killing vacuum for the continuous system corresponds to
the ground state of the Hamiltonian $H_0$ in this system. Hence
its density matrix is given by
%============< EQUATION >==============%
%
\begin{equation}
\rho (\{q^A\}, \{{q'}^B \})
=\left(\det\frac{W}{\pi}\right)^{1/2}
\exp\left[ -\frac{1}{2}W_{AB}(q^Aq^B+ {q'}^A {q'}^B)\right]\ \ ,
\label{eqn:rho-vac}
\end{equation}
%======================================%
where $W=(aV)^{1/2}$. In accordance  with the splitting of  
$\{ q^A\}$ into $\{ q^a\}$ and $\{ q^\alpha \}$, the matrices 
$V$, $W$ and 
its inverse $W^{-1}$ naturally 
split into four blocks as 
%============< EQUATION >==============%
%
\begin{eqnarray}
 \left( V\right)_{AB} &=&  \left(
        \begin{array}{cc}
  	V^{(1)}_{ab}	&	\left({V_{int}}\right)_{a\beta}  \\
	({V_{int}}^{T})_{\alpha b} & V^{(2)}_{\alpha\beta}
        \end{array}	\right),  \nonumber \\
 \left( W \right)_{AB} & = & \left(
 \begin{array}{cc}
        A_{ab}                  & B_{a \beta} \\
        (B^{T})_{\alpha b}      & D_{\alpha \beta} 
 \end{array}   \right), 
   \label{eqn:VWblock}                  \\
 \left( W^{-1} \right)_{AB} & = & \left(
 \begin{array}{cc}
        \tilde{A}^{ab}                  & \tilde{B}^{a \beta} \\
        (\tilde{B}^T)^{\alpha b}        & \tilde{D}^{\alpha \beta}	
 \end{array}      \right). \nonumber
\end{eqnarray}
%======================================%
Taking the partial trace of $\rho (\{q^A\}, \{{q'}^B \})$
 for the inside modes $\{ q^a\}$ $(a=1,\cdots,n_B)$, we get 
 the reduced density matrix 
%============< EQUATION >==============%
%
\begin{eqnarray*}
\rho_{\rm red} (\{q^\alpha \}, \{{q'}^\beta \})
    & = &\left(\det \pi \tilde{D} \right)^{-1/2}
       \exp\left[- \frac{1}{2}(\tilde{D}^{-1})_{\alpha \beta}
        (q^\alpha q^\beta + {q'}^\alpha  {q'}^\beta )  \right. \\
    \qquad & &\left.  -\frac{1}{4} (B^T A^{-1} B)_{\alpha \beta} 
      (q-q')^\alpha (q-q')^\beta \right].
\end{eqnarray*}
%======================================%

Now  the entanglement entropy associated with 
 the boundary $\Sigma$, $S_{ent}:=-{\rm Tr}\rho_{\rm red}\ln \rho_{\rm
red}$, is given as follows\cite{BKLS,Sred}. 
Let $\{ \lambda_i \}$ $(i=1,\cdots,N-n_B)$
 be the eigenvalues of a positive definite 
symmetric matrix\footnote{
 The corresponding expression in ref.\cite{BKLS} 
 (``${\Lambda^a}_b:= (M^{-1})^{ac}N_{cb}$") 
 reads $\Lambda=\tilde{D} B^T A^{-1} B$ 
in the present notation. This definition 
does not give a symmetric matrix and  
should be replaced by 
``$\Lambda^{ab}:= (M^{-1/2})^{ac}N_{cd} (M^{-1/2})^{db}$",
 namely  Eq.(\ref{eqn:Lambda}).
} 
$\Lambda$,
%============< EQUATION >==============%
%
\begin{equation}
\Lambda:= \tilde{D}^{1/2} B^T A^{-1} B \tilde{D}^{1/2}.
\label{eqn:Lambda}
\end{equation}
%======================================%
Then it is easily shown that  modes labeled
by $(l,m)$  contribute to the entangle entropy 
by the amount 
%============< EQUATION >==============%
%
\begin{eqnarray}
    S_{ent}^{(l)} &=& \sum_{i=1}^{N-n_B} S_i , \nonumber \\
   S_i &=& -\frac{\mu_{i}}{1-\mu_{i}}\ln{\mu_{i}} 
        - \ln (1-\mu_{i}),
\label{eqn:Sl}
\end{eqnarray}
%======================================%
where 
$\mu_i := \lambda_i^{-1} \left( \sqrt{1+\lambda_i} -1\right)^2$. 
(Note that $0 < \mu_i < 1$.) Clearly $S_{ent}^{(l)}$ 
is independent of    $m=(-l,-l+1,\cdots,l-1,l)$. 
The entanglement entropy is given by 
%============< EQUATION >==============%
%
\begin{equation}
        S_{ent} = \sum_{l=0}^{\infty}(2l+1)S_{ent}^{(l)}.
        \label{eqn:Sent}
\end{equation}
%======================================%

From Eq.(\ref{eqn:Vlm}), one can  easily show that 
%============< EQUATION >==============%
%
\[
    S_{ent}^{(l)} \sim O\left( (la/r_0)^{-4}\ln (la/r_0)\right)\ \ 
    {\rm as}\ \ la/r_0 \to \infty .
\]
%======================================%
Thus the infinite series Eq.(\ref{eqn:Sl}) actually converge
so that we can safely truncate them 
at some appropriate $l$, depending on the accuracy we require
and the ratio $r_0/a$ we set.

\subsection{Numerical estimation}

Using these formulas, we have evaluated $S_{ent}$ numerically and have
examined its dependence on the area of the boundary, ${\cal A}=4\pi
r_B^2$. In this calculation the outer numerical boundary is set at
$N=60$. The summation in $l$ in Eq.(\ref{eqn:Sent}) is taken has up to
$l=\left[ 10r_0/a \right]$ ($\left[\ \ \right]$ is the Gauss symbol).
From the above asymptotic behavior of $S^{(l)}_{ent}$, this guarantees
the accuracy of $10 \%$.

The result is shown in {\it Figure} \ref{fig:Sent}.
From this figure we see that $S_{ent}$ is proportional to  ${\cal A}/a^2$ 
if we change $r_0$ with fixed $n_B$, and its coefficient only has
a weak dependence on $n_B$. In particular, for the limit
$n_B=1$,  we get 
\begin{equation}
     S_{ent} \simeq  0.024 {\cal A}/a^2 \ \ .
\label{eqn:resultS}
\end{equation}

This result is essentially the same as our previous result for
models in a flat spacetime including the numerical coefficient\cite{paper1}. 
This can be understood in the following way.

Let us make a  coordinate change from $r$ to $x$ defined by  
\[
\frac{r}{r_0}= \frac{(x+1)^2}{4x} ,
\] 
or 
\[
x= \frac{2r}{r_0}-1 +\sqrt{\left(\frac{2r}{r_0}-1\right)^2 - 1} .
\] 
Then Eq.(\ref{eqn:usualmetric}) is rewritten as
\[
ds^2 =   -\left(\frac{x-1}{x+1}\right)^2 dt^2 +
                   r_0^2 \left(\frac{1+x}{2x}\right)^4 
                     (dx^2 + x^2 d\Omega^2) .
\]
Note that $r=0$, $r_0$ and $\infty$ 
correspond to $x=0$, $1$ and $\infty$, respectively. 
It is easy to see that the Hamiltonian 
(Eq.(\ref{eqn:Hamiltonian1})) is given in this coordinate system
as
\begin{eqnarray}
        &H  =& \sum_{lm}H_{lm} ,   \nonumber \\
        & H_{lm} &= \int d\xi
        \frac{64 x^4 (x-1)}{(x+1)^7}
      \left[
        \frac{1}{2}P_{lm}^2
           + \frac{1}{2} \left(\frac{(x+1)^2}{4x} \right)^4 
                     \left\{ (\partial_\xi \varphi_{lm})^2 
                     +\frac{l(l+1)}{\xi^2} \varphi_{lm}^2
                      \right\} \right]  ,
\label{eqn:H-note}         
\end{eqnarray}
where $\xi:=r_0 x$, and $P_{lm}$ and $\varphi_{lm}$ are expressed as
$P_{lm}:=\frac{r_0}{64}\frac{(x+1)^7}{x^4(x-1)} \dot{\phi}_{lm}$ and
$\varphi_{lm}:=r_0 \phi_{lm}$ in terms of $\phi_{lm}$ in
Eq.(\ref{eqn:expansion}).  Here note that the vacuum state is only
weakly dependent on the prefactor $ \frac{64 x^4 (x-1)}{(x+1)^7}$ in
Eq.({\ref{eqn:H-note}}).  If we neglect this prefactor, the vacuum
state is determined by the Hamiltonian which coincides with that for
the flat spacetime at $x=1$. On the other hand, $S_{ent}$ depends on
the modes in a thin layer around the boundary $\Sigma$, whose typical
thickness is a few times of $a\simeq l_{\rm Pl}$). Therefore, when
$\Sigma$ is near the horizon, the value of $S_{ent}$ should be well
approximated the flat spacetime value.

%%%%%%%%%%%%%%%%%%%%%%%%%%%%%%%%%%%%%%%%%%%%%%%%%%%%%%%%%%%%%%%%%
%%%%%%%%%%%%%%%%%%%%%%%%%%%%%%%%%%%%%%%%%%%%%%%%%%%%%%%%%%%%%%%%%
%SECTION 4
%%%%%%%%%%%%%%%%%%%%%%%%%%%%%%%%%%%%%%%%%%%%%%%%%%%%%%%%%%%%%%%%%
%%%%%%%%%%%%%%%%%%%%%%%%%%%%%%%%%%%%%%%%%%%%%%%%%%%%%%%%%%%%%%%%%
		
\section{Entanglement energy}
\label{sec:energy}

In this section we give formulas for the various definitions
of the entanglement energy introduced in  Sec.\ref{subsec:energy},
and estimate their values numerically.

First we derive formulas for the entanglement energies corresponding
to the definitions (I) and (I'):
%============< EQUATION >==============%
%
\begin{eqnarray}
&& E_{ent}^{(I)} :=  \langle :H_{in}:\rangle,
\label{eqn:E-I}\\
&& E_{ent}^{(I')}:=  \langle :H_{out}:\rangle.
\label{eqn:E-I'}
\end{eqnarray}
%======================================%

By  rescaling   the variables $\left\{ q^A\right\}$  in 
\S\ref{sec:entropy} as
%============< EQUATION >==============%
%
\[
 \bar{q}^A := \delta^{AB}\left( W^{1/2}\right)_{BC}q^C,
\]
%======================================%
the expression of 
the density matrix for the vacuum state Eq.(\ref{eqn:rho-vac}) 
gets simplified as 
%============< EQUATION >==============%
%
\[
 \langle\left\{\bar{q}^A\right\}|\rho|\left\{\bar{q}'^B\right\}\rangle 
	= \prod_{C=1}^N\pi^{-1/2}\exp\left[ -\frac{1}{2}\left\{
		(\bar{q}^C)^2 + (\bar{q}'^C)^2\right\}\right],
\]
%======================================%
and the normal ordered Hamiltonian $:H_{in}:$  is
represented as
%============< EQUATION >==============%
%
\begin{eqnarray*}
 :H_{in}: & = & -\frac{1}{2a}\delta^{ab}
	\left(\frac{\partial}{\partial q^a}-w_{ac}^{(1)}q^c\right)
	\left(\frac{\partial}{\partial q^b}+w_{bd}^{(1)}q^d\right)
	\nonumber\\
 & = & -\frac{1}{2a}U^{AB}
	\left(\frac{\partial}{\partial\bar{q}^A}
		-\bar{w}^{(1)}_{AC}\bar{q}^C\right)
	\left(\frac{\partial}{\partial\bar{q}^B}
		+\bar{w}^{(1)}_{BD}\bar{q}^D\right).
\end{eqnarray*}
%======================================%
Here $w^{(1)}$ is the positive square-root of $aV^{(1)}$, 
and the matrices $U$ and $\bar w^{(1)}$ are defined as
%============< EQUATION >==============%
%
\begin{eqnarray*}
 U^{AB} & := & \delta^{AC}\left( W^{1/2}\right)_{Ca}\delta^{ab}
	\left( W^{1/2}\right)_{bD}\delta^{DB},\\
 \bar{w}^{(1)}_{AB} & := & \delta_{AC}\left(W^{-1/2}\right)^{Ca}
	w^{(1)}_{ab}\left(W^{-1/2}\right)^{bD}\delta_{DB}.
\end{eqnarray*}
%======================================%
Hence the matrix elements of $:H_{in}:\rho$ with respect to
the basis $|{\bar q^A}\rangle$ are expressed as
%============< EQUATION >==============%
%
\begin{eqnarray*}
  \langle\left\{\bar{q}^A\right\}| :H_{in}:\rho 
		|\left\{\bar{q}^B\right\}\rangle & = & 
	\frac{1}{2a}\left\{\left[
		(\bar{w}^{(1)}+1)U(\bar{w}^{(1)}-1)
		\right]_{AB}\bar{q}^A\bar{q}^B
	+ {\rm Tr}\left[ U(1-\bar{w}^{(1)})\right]\right\}
			\nonumber\\
 & & \times\prod_{C=1}^N\pi^{-1/2}\exp\left[ -(\bar{q}^C)^2\right].
\end{eqnarray*}
%======================================%
From this we obtain\footnote{
             See Eq.(\ref{eqn:VWblock}) for the definitions of 
             the matrices $A$, $\tilde{A}$, $D$ and $\tilde{D}$.
              }  
%============< EQUATION >==============%
%
\begin{eqnarray}
 E_{ent}^{(I)} & = & \int\left(\prod_{C=1}^N d\bar{q}^C\right)
	\langle\left\{\bar{q}^A\right\}| :H_{in}:\rho 
		|\left\{\bar{q}^B\right\}\rangle \nonumber\\
 & = & \frac{1}{4a}\left[ aV^{(1)}_{ab}(\tilde{A})^{ab}
	+ A_{ab}\delta^{ab} -2w^{(1)}_{ab}\delta^{ab}\right].
	\label{eqn:E-I-formula}
\end{eqnarray}
%======================================%
Similarly $E_{ent}^{(I')}$ is expressed as
%============< EQUATION >==============%
%
\begin{equation}
 E_{ent}^{(I')} = \frac{1}{4a}\left[ 
	aV^{(2)}_{\alpha\beta}(\tilde{D})^{\alpha\beta}
	+ D_{\alpha\beta}\delta^{\alpha\beta} 
	-2w^{(2)}_{\alpha\beta}\delta^{\alpha\beta}\right],
	\label{eqn:E-I'-formula}
\end{equation}
%======================================%
where $w^{(2)}$ is the positive square-root of $aV^{(2)}$.

$E_{ent}$ corresponding to the definition $(III)$ are simply related 
to $E^{(I)}_{ent}$ by
%============< EQUATION >==============%
%
\begin{equation}
 E_{ent}^{(III)} = M + E_{ent}^{(I)} 
\label{eqn:E-III} 
\end{equation}
%======================================%
Further, $E_{ent}$ corresponding to the definitions (IV) 
have already been given for the flat case\cite{paper1}. They are
expressed as
%============< EQUATION >==============%
%
\begin{eqnarray}
	E_{ent}^{(IV_1)} & := & {\rm Tr}\left[:H_{tot}:\rho'\right],
	                          \nonumber\\
	E_{ent}^{(IV_2)} & := & 
		{\rm Tr}\left[ (:H_{in}: + :H_{out}:)\rho\right].
\label{eqn:E-IV}		
\end{eqnarray}
%======================================%
Here  $H_{tot}:=H_{in}+H_{out}+H_{int}$ and  
$\rho':=\rho_{in} \otimes \rho_{out}$\cite{paper1}.\footnote{Note that  
         $E_{ent}^{(IV_1)}$ and  $E_{ent}^{(IV_2)}$ here 
         are the generalizations to the curved background case
         of  $E_{ent}^I$ and $E_{ent}^{II}$, respectively, 
         in Ref.\cite{paper1}. In the same way 
         the subscripts `$in$' and `$out$' correspond 
         to `$1$' and `$2$' in  Ref.\cite{paper1}. 
         }
In particular for  a vacuum state, they become\cite{paper1}  
%============< EQUATION >==============%
%
\begin{eqnarray}
	E_{ent}^{(IV_1)} & = & -\frac{1}{2} {\rm Tr}
		\left[V_{int}^T\tilde{B}\right] \nonumber\\
	E_{ent}^{(IV_2)} & = & -\frac{1}{2} {\rm Tr}
		\left[V_{int}^T\tilde{B}\right] -
		\frac{1}{2a}\left({\rm Tr}\left[(aV^{(1)})^{1/2}\right]
		+ {\rm Tr}\left[(aV^{(2)})^{1/2}\right] 
		- {\rm Tr}\left[W \right]\right),
\label{eqn:E-IV-2}
\end{eqnarray}
%======================================%
where $V_{int}$, $V^{(1,2)}$ and $W$ are given 
in Eq.(\ref{eqn:VWblock}).

Like the entanglement entropy (Eq.(\ref{eqn:Sent})),
 the entanglement energy is also given by the summation  of  
each contribution specified by $(l,m)$:
%============< EQUATION >==============%
%
\begin{equation}
	E_{ent} = \sum_{l=0}^{\infty}(2l+1)E_{ent}^{(l)}.
	\label{eqn:EentL}
\end{equation}
%======================================%
Here $E_{ent}$ represents any kind of the entanglement energy 
defined above (except for $E_{ent}^{(II)}=M$ and 
$E_{ent}^{(III)}=M+E_{ent}^{(I)}$)
  and $E_{ent}^{(l)}$ is its $(l,m)$-contribution. 
 From Eq.(\ref{eqn:Vlm}), it is  easily shown that 
%============< EQUATION >==============%
%
\[
       E_{ent}^{(l)} \sim  O\left( (la/r_0)^{-3}\right)\ \ {\rm as}\ \ 
      la/r_0 \to \infty.
\]
%======================================%
 Thus the summation with respect to $l$ in (\ref{eqn:EentL})
converges.

With the helps of these formulas, we have numerically evaluated
$E_{ent}^{(I)}$, $E_{ent}^{(I')}$ $E_{ent}^{(IV_1)}$ and
$E_{ent}^{(IV_2)}$. Now we have taken the numerical outer boundary at
$N=100$. The truncation in the $l$-summation is the same as for
$S_{ent}$ (up to $l=[10r_0/a]$), which implies that the accuracy is
about $10\%$ from the above asymptotic estimate for $E^{(l)}_{ent}$.

The results of numerical calculations are shown in {\it Figure}
\ref{fig:E-I}, {\it Figure} \ref{fig:E-I'}, {\it Figure}
\ref{fig:E-IV-1} and {\it Figure} \ref{fig:E-IV-2}. In these
figures, $r_B E_{ent}$ is plotted as a function of
$(r_B/a)^2$ for $n_B=1,2,5$. 
All of these figures show that $r_BE_{ent}$ is proportional
to $n_B(r_B/a)^2$: 
%============< EQUATION >==============%
%
\begin{eqnarray}
 E_{ent}^{(I)} & \sim & 0.05 (n_B-1/2)r_B/a^2,\nonumber\\
 E_{ent}^{(I')} & \sim & 0.05 (n_B+1/2)r_B/a^2,\nonumber\\
 E_{ent}^{(IV_1)} & \sim & 0.2n_Br_B/a^2,\nonumber\\
 E_{ent}^{(IV_2)} & \sim & 0.1n_Br_B/a^2.\nonumber\\
\end{eqnarray}
%======================================%
Note that the identity $E_{ent}^{(IV_2)}=E_{ent}^{(I)}+E_{ent}^{(I')}$ 
holds from the definitions.
 
From these equations we immediately see that the values of
$E_{ent}^{(I)}$ and $E_{ent}^{(I')}$ coincide except for a tiny
difference independent of $n_B$. As was mentioned in
Sec.\ref{subsec:energy}, this difference is understood by the
gravitational red-shift:
$E_{ent}^{(I)}$ comes from the modes just
inside $\Sigma$ while $E_{ent}^{(I')}$ originates from the modes just
outside $\Sigma$. In the present numerical calculations, it means that
$E_{ent}^{(I)}$ and $E_{ent}^{(I')}$ are determined by the modes at
$\rho=(n_{B}-1/2)a$ and $\rho=(n_{B}+1/2)a$, respectively(see
Eq.(\ref{eqn:Vlm})).  Hence, taking account of the fact that 
the contribution of each mode to the entanglement energy is proportional to
the red-shift factor at its location, the ratio of $E_{ent}^{(I)}$
and $E_{ent}^{(I')}$ should be approximately given by 
%============< EQUATION >==============%
%
\begin{eqnarray}
 E_{ent}^{(I)}:E_{ent}^{(I')}
 & \sim & \nu (\rho =(n_{B}-1/2)a):\nu (\rho =(n_{B}+1/2)a)\nonumber\\
 & \sim & (n_{B}-1/2):(n_{B}+1/2).
\end{eqnarray}
%======================================%
This is consistent with the above numerical result.

This argument is also supported by the numerical result for the flat
spacetime model shown in {\it Figure} \ref{fig:E-I-Flat}. In this
figure the values of $E_{ent}^{(I)}$ and $E_{ent}^{(I')}$ for a
massless scalar field in the Minkowski spacetime with $\Sigma=S^2$ are
plotted. In this case there is no gravitational red-shift effect, so
we expect that $E_{ent}^{(I)} = E_{ent}^{(I')}$, as confirmed by
the numerical calculation.

%%%%%%%%%%%%%%%%%%%%%%%%%%%%%%%%%%%%%%%%%%%%%%%%%%%%%%%%%%%%%%%%%
%%%%%%%%%%%%%%%%%%%%%%%%%%%%%%%%%%%%%%%%%%%%%%%%%%%%%%%%%%%%%%%%%
%SECTION 5
%%%%%%%%%%%%%%%%%%%%%%%%%%%%%%%%%%%%%%%%%%%%%%%%%%%%%%%%%%%%%%%%%
%%%%%%%%%%%%%%%%%%%%%%%%%%%%%%%%%%%%%%%%%%%%%%%%%%%%%%%%%%%%%%%%%

\section{Comparison between entanglement thermodynamics  and
black-hole thermodynamics}
\label{sec:discussions}

Now on the basis of  our results for $S_{ent}$ and $E_{ent}$, 
let us  compare the structure of the entanglement thermodynamics with
that of black hole.
  
From the numerical results in \S3 and \S4, the entanglement entropy
$S_{ent}$ and the entanglement energy $E_{ent}$ for our model
are expressed as
%============< EQUATION >==============%
%
\begin{eqnarray}
&& S_{ent}=C_S \left(\frac{l_{\rm Pl}}{a} \right)^2
           \frac{{\cal A}}{4l_{\rm Pl}^2} \quad 
            (C_S \simeq 0.096), \\
\label{eqn:summaryS}
&& E_{ent} =  C_E  {r_B \over 2a^2} 
         = C_E \left(\frac{r_B}{r_0}\right)
              \left(\frac{l_{\rm Pl}}{a} \right)^2 M c^2,
\label{eqn:summaryE}
\end{eqnarray}
%======================================%
where
%============< EQUATION >==============%
\begin{eqnarray}
C_{E}^{(I)} &\simeq& 0.05, \ \ 
C_{E}^{(I')} \simeq 0.15, \nonumber \\
C_{E}^{(II)} &=& \left(\frac{r_0}{r_B}\right)
              \left(\frac{a}{l_{\rm Pl}} \right)^2 
              \simeq \left(\frac{a}{l_{\rm Pl}} \right)^2, 
                                 \nonumber \\
C_{E}^{(III)} &=& \left(\frac{r_0}{r_B}\right)
              \left(\frac{a}{l_{\rm Pl}} \right)^2
               + C_{E^{(I)}} \simeq
              \left(\frac{a}{l_{\rm Pl}} \right)^2
               + C_{E^{(I)}}, \nonumber \\
C_{E}^{(IV_1)} &\simeq& 0.4, \ \ 
     C_{E}^{(IV_2)} \simeq 0.2.
\end{eqnarray}
%
%======================================%
Here and hereafter we will only consider the horizon limit, $n_B =1$. 

It is helpful  to keep in mind  that   
 the case $C_S=C_E=1$ along with $a=l_{\rm Pl}$ 
corresponds to the black hole thermodynamics.\footnote{
Strictly speaking $\cal A$ in Eq.(\ref{eqn:summaryS}) is the 
area of the boundary and it differs from the area of the horizon.
However we here ignore  the
 difference since it is totally  negligible  as 
 compared with $\cal A$ itself.  Similarly 
 $\frac{r_B-r_0}{r_0}=O\left((\frac{a}{r_0})^2\right)$ so that 
 we can safely set $r_B\simeq r_0$ in the present context. 
 }  

From these expressions and the first law of thermodynamics
%============< EQUATION >==============%
%
\begin{equation}
dE=TdS
\label{eqn:1stlaw'}
\end{equation}
%
%======================================%
the entanglement temperature $T_{ent}$ is determined as
%============< EQUATION >==============%
%
\begin{equation}
T_{ent}={C_E\over C_S}  T_{BH}.
\end{equation}
%
%======================================%
Thus we get 
%============< EQUATION >==============%
\begin{eqnarray}
T_{ent}^{(I)} &\simeq& 0.52 T_{BH}, \ \ 
  T_{ent}^{(I')} \simeq 1.6  T_{BH} \ \ , \nonumber \\
T_{ent}^{(II)} &\simeq& \left(\frac{a}{l_{\rm Pl}} \right)^2 T_{BH} \ \ , 
                                 \nonumber \\
T_{ent}^{(III)} &\simeq& 
        \left\{ \left(\frac{a}{l_{\rm Pl}} \right)^2
        + 0.52 \right\} T_{BH} \ \ , \nonumber \\
T_{ent}^{(IV_1)} &\simeq& 4.2 T_{BH}, \ \  
     T_{ent}^{(IV_2)} \simeq 2.1 T_{BH}.
\end{eqnarray}
%
%======================================%

These results have several interesting features. First of all we
immediately see that the entanglement thermodynamics on the
Schwarzschild spacetime show exactly the same behavior as the black
hole thermodynamics, as summarized in {\it Table} 
\ref{table:comparison}. This behavior is just what we expected
from the intuitive argument in the introduction: the gravitational
redshift effect modifies the area dependence of $E_{ent}$ so as to
make the entanglement thermodynamics behave just like the black hole
thermodynamics.

Second it should be noted that the temperature $T_{ent}$ becomes
independent of the cut-off scale $a$ only for the options type $(I)$
and $(IV)$ (i.e.  $(I)$, $(I')$, $(IV_1)$ and $(IV_2)$).  This
indicates that (II) and (III) are not good definitions of the energy
in our approach. It is also suggestive that the 
average $\frac{1}{2}(T_{ent}^{(I)}+T_{ent}^{(I')})$ gives almost the
same value as $T_{BH}$. This averaging corresponds to averaging out
the difference in the red-shift factors for the one-mesh `inside' and
the one-mesh `outside' of the boundary.  Therefore such an averaging
may have some meaning.

To summarize, our model analysis strongly suggests a tight 
connection between the entanglement thermodynamics and the black 
hole thermodynamics. It is worth emphasizing that matter fields on 
a black hole spacetime capture the background characteristics. 
Of course, our model is too simple to give any
definite conclusion based on it. In particular, the ambiguity in
the definition of the energy comes from neglecting
backreaction of the quantum field on gravity. Further,
even in the fixed background framework, our model is too simple in
that its thermodynamics is essentially controlled by one parameter
corresponding to the black hole mass. It is obviously useful to
see whether the entanglement thermodynamics is consistent with
the black hole thermodynamics for models with more parameters, 
such as those on the Reisner-Nordstrom spacetime or on the Kerr 
spacetime before attacking the difficult task of going beyond the 
semi-classical approximation.

\section*{Acknowledgments} 
The authors thank T. Jacobson for a valuable comment on the energy
red-shift in a black hole spacetime.  Numerical computation in this
work was carried out at the Yukawa Institute Computer Facility. One of
the author (HK) is supported by the Grant-In-Aid of for Scientific
Research (C) of the Ministry of Education, Science, Sports and Culture 
in Japan(05640340).

%======================================%
%<<<<<<<<<<<< REFERENCES >>>>>>>>>>>>>>%
%======================================%

\newpage

%============< FIGURE >==============%
%              Kruskal.epsf
\begin{figure}
\caption{
The Kruskal extension of the Schwarzschild spacetime. We
consider only the region $I$ (the shaded region). As the boundary
$\Sigma$ we take the hypersurface $r=r_B$. 
}
\label{fig:Kruskal}
\end{figure}
%======================================%

%============< FIGURE >==============%
%              Sent.epsf
\begin{figure}
\caption{
The numerical evaluations for $S_{ent}$ for the discretized theory of
the scalar field in Schwarzschild spacetime. $S_{ent}$ for $n_B=1,2,5$
is shown as functions of $A/a^2$, where $A$ is area of the boundary. 
We have taken $N=100$ and performed the summation over $l$ up to
$10r_0/a$. 
}
\label{fig:Sent}
\end{figure}
%======================================%

%============< FIGURE >==============%
%              E-I.epsf
\begin{figure}
\caption{
The numerical evaluations for $E_{ent}^{(I)}$ for the
discretized theory of the scalar field in Schwarzschild
spacetime. $r_BE_{ent}^{(I)}$ for $n_B=1,2,5$ is shown as
functions of $(r_B/a)^2$, where $r_B\equiv r(\rho =n_Ba)$. We have
taken $N=100$ and performed the summation over $l$ up to $10r_0/a$. 
}
\label{fig:E-I}
\end{figure}
%======================================%

%============< FIGURE >==============%
%              E-I-.epsf
\begin{figure}
\caption{
The numerical evaluations for $E_{ent}^{(I')}$ for the
discretized theory of the scalar field in Schwarzschild
spacetime. $r_BE_{ent}^{(I')}$ for $n_B=1,2,5$ is shown as
functions of $(r_B/a)^2$, where $r_B\equiv r(\rho =n_Ba)$. We have
taken $N=100$ and performed the summation over $l$ up to $10r_0/a$. 
}
\label{fig:E-I'}
\end{figure}
%======================================%

%============< FIGURE >==============%
%              E-IV-1.epsf
\begin{figure}
\caption{
The numerical evaluations for $E_{ent}^{(IV_1)}$ for the discretized
theory of the scalar field in Schwarzschild
spacetime. $r_BE_{ent}^{(IV_1)}$ for $n_B=1,2,5$ is shown as functions of
$(r_B/a)^2$, where $r_B\equiv r(\rho =n_Ba)$. We have taken $N=100$
and performed the summation over $l$ up to $10r_0/a$. 
}
\label{fig:E-IV-1}
\end{figure}
%======================================%

%============< FIGURE >==============%
%              E-IV-2.epsf
\begin{figure}
\caption{
The numerical evaluations for $E_{ent}^{(IV_2)}$.
 $r_BE_{ent}^{(IV_2)}$ for $n_B=1,2,5$ is shown as
functions of $(r_B/a)^2$, where $r_B\equiv r(\rho =n_Ba)$. We have
taken $N=100$ and performed the summation over $l$ up to $10r_0/a$. 
}
\label{fig:E-IV-2}
\end{figure}
%======================================%

%============< FIGURE >==============%
%              E-I-Flat.epsf
\begin{figure}
\caption{
The numerical evaluations for $E_{ent}^{(I)}$ and $E_{ent}^{(I')}$ 
in the Minkowski spacetime.
 $aE_{ent}^{(I)}$ and $aE_{ent}^{(I')}$ are shown 
as functions of $(r_B/a)^2$, where $r_B$ is the radius of the sphere 
which provides the division of region. We have taken $N=60$ and performed 
the summation over $l$ up to $10r_0/a$. 
}
\label{fig:E-I-Flat}
\end{figure}
%======================================%

\newpage

%============< TABLE >==============%
%             
\begin{table}
\caption{Comparison of two kinds of thermodynamics}
\label{table:summary}
 \begin{center}
  \begin{tabular}{rcc} \hline
                & {\it Entanglement in Schwarzschild spacetime}  
                & {\it Black-Hole} \\ \hline
   {\it Varied} & $A$                   & $A$ \\
   {\it Fixed}  & $a$                   & $l_{\rm Pl}$ \\
        $S$     & $\propto A$           & $\propto A$ \\
        $E$     & $\propto A^{1/2}$     & $\propto A^{1/2}$ \\
        $T$     & $\propto A^{-1/2}$    & $\propto A^{-1/2}$ \\ \hline
   {\it Varied} & $a$                   & $l_{\rm Pl}$ \\
   {\it Fixed}  & $A$                   & $A$ \\
        $S$     & $\propto a^{-2}$      & $\propto l_{\rm Pl}^{-2}$ \\
        $E$     & $\propto a^{-2}$      & $\propto l_{\rm Pl}^{-2}$ \\
        $T$     & $\propto a^{0}$      & $\propto l_{\rm Pl}^{0}$ \\ 
                                                                \hline 
   {\it Varied} & $a$                   & $l_{\rm Pl}$ \\
   {\it Fixed}  & $E_{ent}$             & $M$ \\
        $S$     & $\propto a^2$           & $\propto l_{\rm Pl}^2$ \\
        $E$     & $\propto a^0$         & $\propto l_{\rm Pl}^0$ \\
        $T$     & $\propto a^{-2}$      & $\propto l_{\rm Pl}^{-2}$ \\ \hline
    \end{tabular}
 \end{center}
\label{table:comparison}
\end{table}
%======================================%

\end{document}